\documentclass[journal]{IEEEtran}

\usepackage{amsmath,amsfonts,amssymb}
\usepackage{graphicx}
\usepackage[colorlinks=true, allcolors=blue]{hyperref}

\usepackage{subfigure}
\usepackage{multirow}
\usepackage{gensymb}
\usepackage{mathtools}
\newcommand{\specialcell}[2][c]{\begin{tabular}[#1]{@{}c@{}}#2\end{tabular}}

\begin{document}

\title{Do Noises Bother Human and Neural Networks In the Same Way? A Medical Image Analysis Perspective}

\author{
Shao-Cheng Wen$^{1}$,
Yu-Jen Chen$^{1}$, 
Zihao Liu$^{2}$,
Wujie Wen$^{3}$,
Xiaowei Xu$^{4}$, \\
Yiyu Shi$^{5}$,
Tsung-Yi Ho$^{1}$,
Qianjun Jia$^{4}$, 
Meiping Huang$^{4}$,
Jian Zhuang$^{4}$
\\
\textit{$^{1}$Department of Computer Science, National Tsing Hua University, Hsinchu, Taiwan\\ 
$^{2}$Department of Electrical and Computer Engineering, Florida International University, FL, USA\\
$^{3}$Department of Electrical and Computer Engineering, Lehigh University, PA, USA\\
$^{4}$Guangdong Provincial People's Hospital, Guangdong Acadamic of Medical Science, Guangzhou, China\\
$^{5}$Department of Computer Science and Engineering, University of Notre Dame, IN, USA}

\thanks{This research was approved by the Research Ethics Committee of Guangdong General Hospital, Guangdong Academy of Medical Science with the protocol No. 20140316.}
}


%

\maketitle

\begin{abstract}
Deep learning had already demonstrated its power in medical images, including denoising, classification, segmentation, etc. All these applications are proposed to automatically analyze medical images beforehand, which brings more information to radiologists during clinical assessment for accuracy improvement. Recently, many medical denoising methods had shown their significant artifact reduction result and noise removal both quantitatively and qualitatively. However, those existing methods are developed around human-vision, i.e., they are designed to minimize the noise effect that can be perceived by human eyes. In this paper, we introduce an application-guided denoising framework, which focuses on denoising for the following neural networks. In our experiments, we apply the proposed framework to different datasets, models, and use cases. Experimental results show that our proposed framework can achieve a better result than human-vision denoising network.

\end{abstract}


\IEEEpeerreviewmaketitle

\section{Introduction}
\label{sec:intro}
\begin{figure}[ht]
\centering
\subfigure[Noise-affected examples]{
\centering
\includegraphics[width=0.45\linewidth]{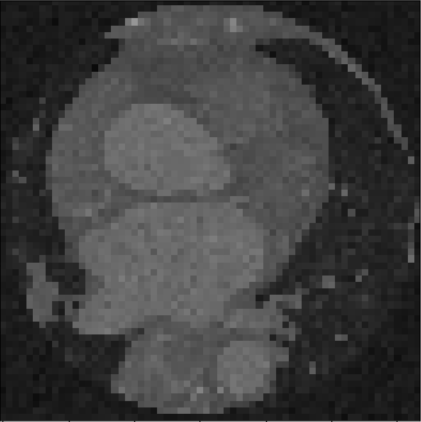}
\includegraphics[width=0.45\linewidth]{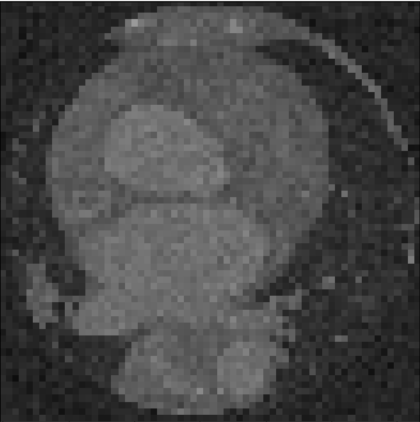}
}\\
\subfigure[Accuracy/Dice vs Noise level]{
\includegraphics[width=0.9\linewidth]{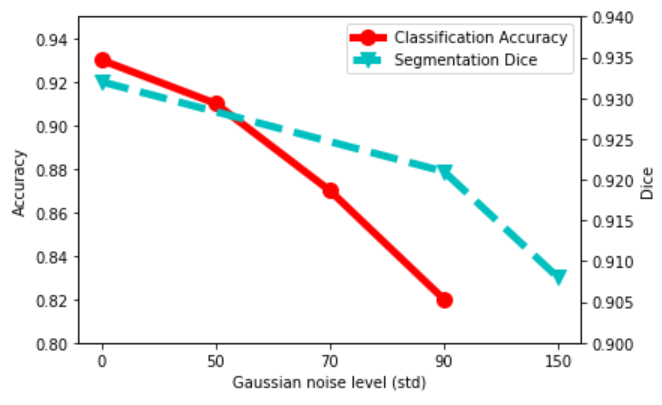}}

\caption{(a) Demonstrates the noise-affected test images with $\mu=0$, $\sigma= 90$ (left) and $\mu=0$, $\sigma= 150$ (right), respectively. (b) segmentation and classification Dice/Accuracy w.r.t. Gaussian noise level. Note that we use the dirty Multi-Modality Whole Heart Segmentation (MM-WHS) dataset to train the segmentation model No-New-Net \cite{isensee2018no} and Classification Convolutional Neural Network (CCNN) \cite{sultan2019multi}. Detailed experimental settings can be found in Section \ref{sec:exp}.}
\label{Fig.noise_level}
\end{figure}

\IEEEPARstart{T}{he} prevalence of deep learning in medical image computing and analysis has greatly reduced the human effort and enhanced the efficiency of diagnosis and treatment \cite{xu2018scaling,ding2020uncertainty,xu2018quantization,wang2020ica}. To achieve superb performance in such tasks, high-quality medical images are often indispensable for training and testing state-of-the-art deep learning models \cite{xu2019whole,xu2020imagechd}. Unfortunately, these raw images inevitably suffer from high-intensity noises (see Fig.~\ref{Fig.noise_level} (a)) due to complex clinical scenarios \cite{boas2012ct,krupa2015artifacts,liu2020multi}, 
which significantly jeopardizing the capability of machine learning models on image segmentation and classification. As the example in Fig.~\ref{Fig.noise_level} (b) shows, both image segmentation performance (Dice) and classification accuracy drop dramatically with the increase of noise level. Even neural network models are trained for better generalization by using images containing the same level of noise as that of testing ones, the performance can be decreased. In contrast, testing images in the medical domain usually can be much noisier than the training dataset. This further aggravates the accuracy problem for deep learning assisted medical imaging.

\begin{figure*}[h]
\centering
\includegraphics[width=0.94\linewidth]{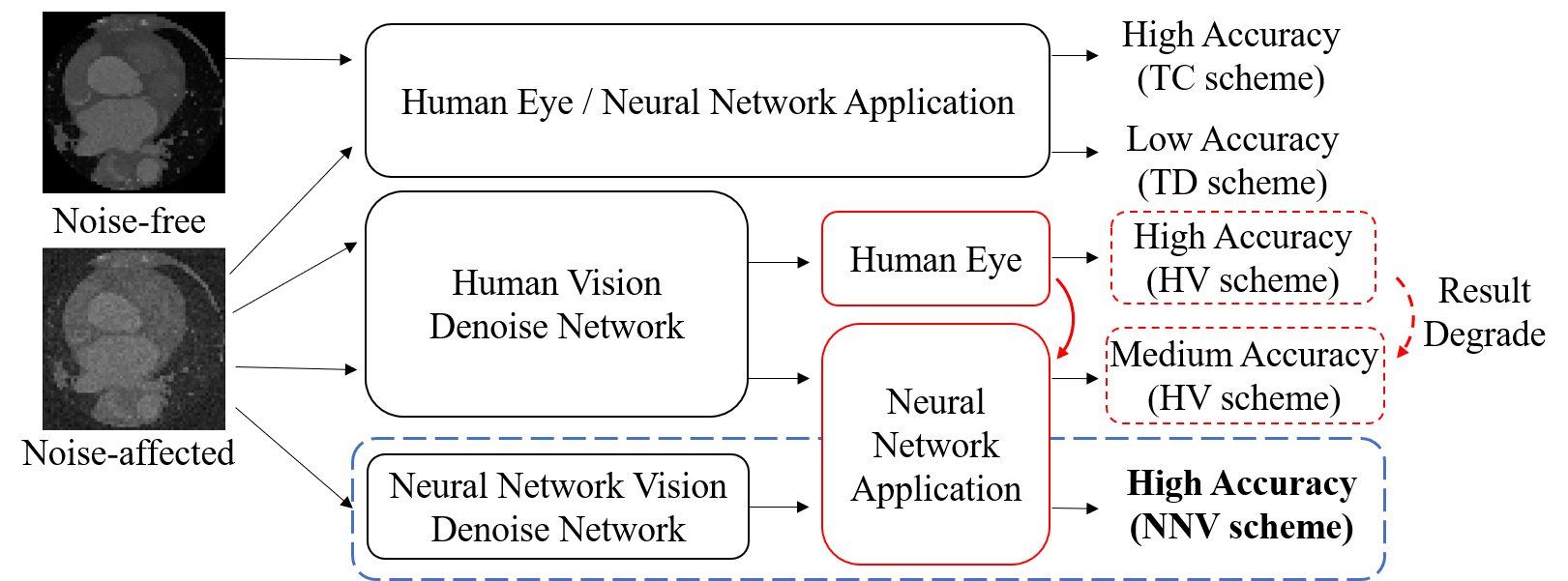}
\caption{Workflow comparison for input images with different noise level and different predict source (human and neural network application). The detail of all four schemes will be introduced in section \ref{sec:exp}. We would like to point out that the result of the human-vision scheme may be degraded as the predict target changed (block marked with red). Note that the blue rectangle marked in the last row is the proposed workflow.}
\label{Fig.workflow}
\end{figure*}

To tackle this issue, image denoising is typically introduced as a pre-processing step before neural-network-based image classification or segmentation. Recently, most existing denoising methods \cite{chen2020zero}, such as Residual Encoder-Decoder Convolution Neural Network (RED-CNN) \cite{chen2017low} and Multi-Channel Denoising Convolution Neural Network (MCDnCNN) \cite{jiang2018denoising} utilized the power of deep neural network, which attempt to learn the distributions of the noise, so as to eliminate the noises in a more elaborate manner.

\begin{figure*}[t]
\centering
\subfigure[Dirty Image]{
\includegraphics[width=0.26\linewidth]{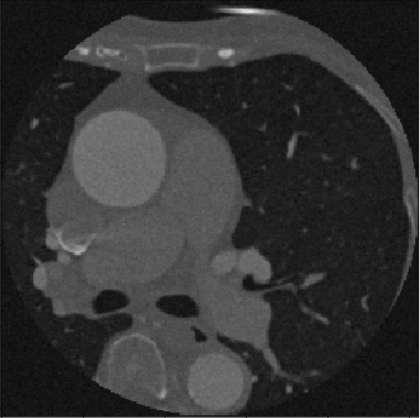}}
\subfigure[HV]{
\includegraphics[width=0.26\linewidth]{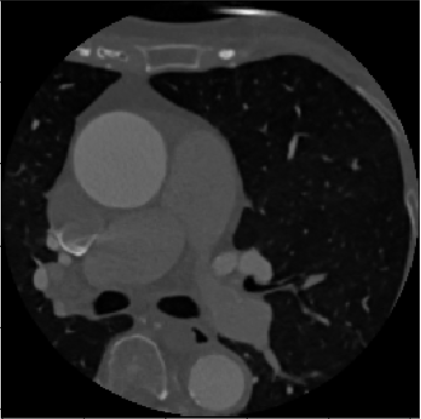}}
\subfigure[NNV]{
\includegraphics[width=0.26\linewidth]{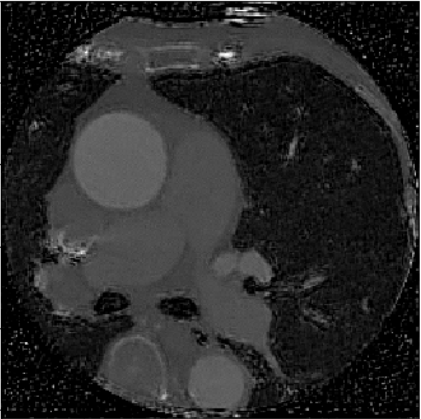}}\\
\subfigure[Ground-truth]{
\includegraphics[width=0.26\linewidth]{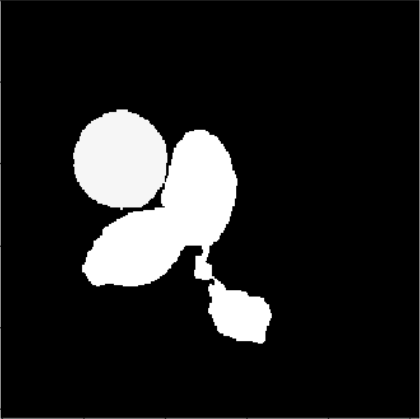}}
\subfigure[Result for (b)]{
\includegraphics[width=0.26\linewidth]{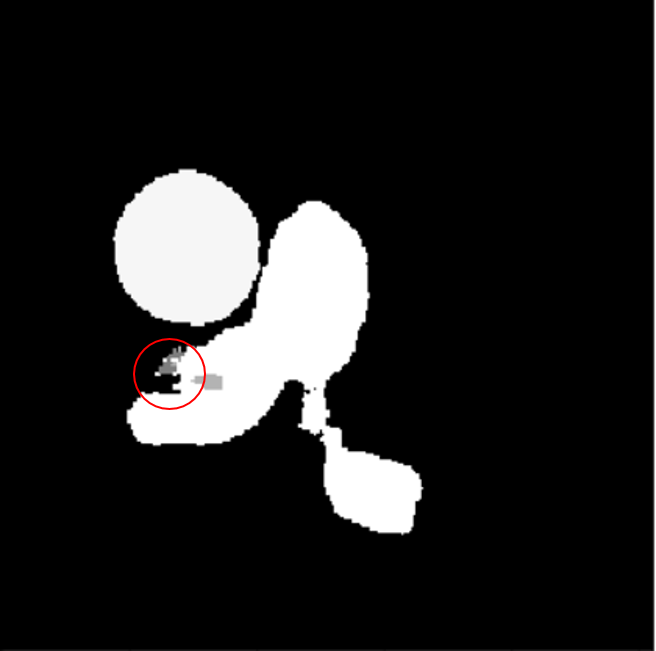}}
\subfigure[Result for (c)]{
\includegraphics[width=0.26\linewidth]{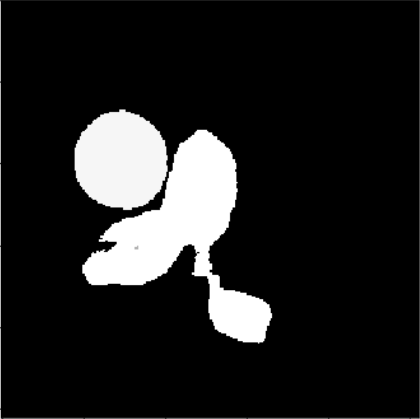}}
\caption{Segmentation result comparison for (a) dirty image and the denoised image through (b) Human-Vision (HV) and (c) Neural-Network-Vision (NNV). The corresponding dice score are 1.0 and 0.704, and 0.849. As shown in (d), the ground-truth segment the image into two class, ascending aorta and the pulmonary artery. However, the region circled with red in (e) is the result of HV misclassified the pulmonary artery into the left atrium blood cavity, while (f) is correctly classified.}
\label{Fig.seg_dice_example}
\end{figure*}

These denoising techniques could largely remove the noises based on the image visual quality measurements defined by human eyes, e.g., peak signal-to-noise ratio (PSNR) calculated by pixel-by-pixel difference between clean and its dirty version, and thereby enhance neural-network-based medical image segmentation or classification performance. We would like to argue that their advocated high denoising efficiency (dedicated to ``human-vision") may not be necessarily translated into impressive accuracy improvement for neural networks (or ``neural-network-vision"). A clear workflows comparison for input images with different noise level and predict environment (human eye or neural network application) could be found in Fig. \ref{Fig.workflow}. From this figure, we would like to point out that the result of the human-vision scheme may be degraded as the predict environment changed.

In this paper, we propose to redefine the framework of medical image denoising by integrating the concept of ``neural-network-vision". Different from the human perceived visual distortion adopted by existing denoising solutions, the proposed framework evaluates the denoising efficiency directly through the perspective of neural network computation. As a result, such denoising can deal with the noises in a way that neural network favors \cite{liu2020connecting,fan2018segmentation}, so as to significantly boost the accuracy. We validate and compare our design with state-of-the-art denoising solutions, through comprehensive experiments on both image segmentation and classification tasks. Segmentation evaluation included two popular segmentation networks, No-New-Net \cite{isensee2018no} 2D and 3D version, under two different datasets, Multi-Modality Whole Heart Segmentation (MM-WHS) and Brain Tumor Segmentation challenge (BraTS). Classification evaluation was conducted through Classification Convolutional Neural Network (CCNN) \cite{sultan2019multi} using Brain Tumor dataset. Examples demonstrated in Fig.~\ref{Fig.seg_dice_example} have shown the qualitative effect of our method on the segmentation results.

The main contributions of our work are as follows:
\begin{itemize}
    \item A novel denoising framework guided by neural-network-vision is proposed.
    \item We proposed the very first application guided denoising network for image denoising by implementing the concept of ``neural-network-vision", and the denoising network can denoise images in a way desirable by any application network.
    \item Experimental results show that the proposed Neural-Network-Vision-based (NNV) image-denoising method outperforms any existing Human-Vision-based (HV) image-denoising methods in both segmentation and classification tasks.
\end{itemize}

\section{Motivation}
\label{sec:motivation}

In this section, we will demonstrate that human eyes and neural networks have very different understandings on image noises:
\begin{enumerate}
    \item Keeping the noises that cannot be perceived by HV are unable to be effectively removed by current denoising methods due to HV-based quality judgement, may lead to considerable accuracy drop on deep-learning-based medical imaging.
    \item Removing all noises that are obvious to human eyes via existing denoising methods according to human-visual rules, in contrast, may degrade the performance of deep-learning-based medical imaging.
\end{enumerate}

\textit{The first argument} has been proved by adversarial examples in deep learning security research--images containing very small and human-imperceptible noises can mislead the decision of deep learning models with high confidence~\cite{szegedy2013intriguing}.

\textit{The second argument} can be roughly interpreted as follows: simply removing all noises perceivable by human eyes to make images completely noise-free, sometimes may result in inferior neural network decision making. Some noises could reinforce the information deemed to be important by neural networks for better image segmentation and recognition.

\begin{figure*}[t]
\centering
 \includegraphics[width=0.9\textwidth]{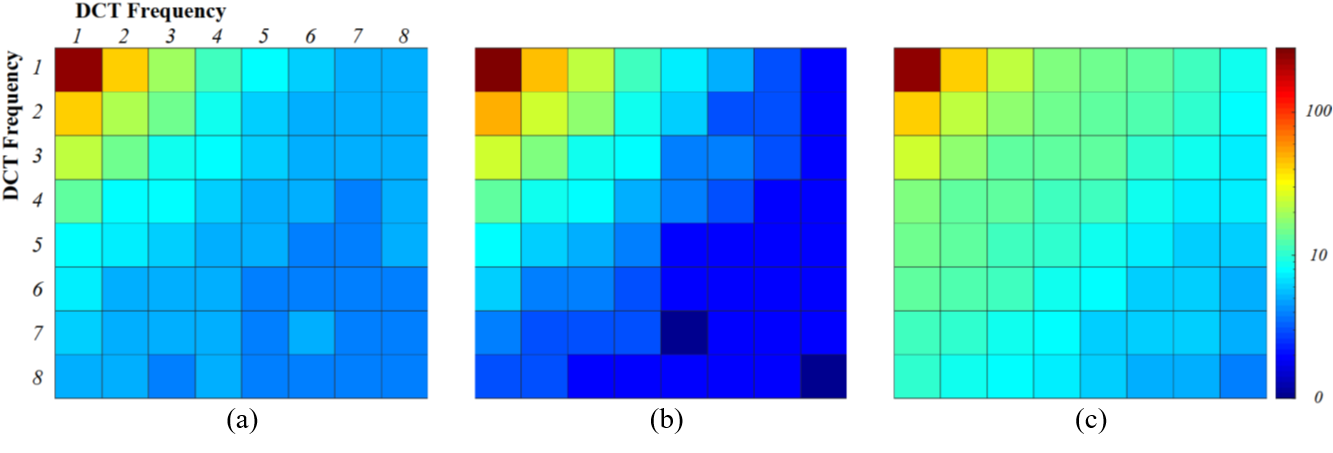}
 \caption{DCT frequency distributions at different examples: (a) Test image; (b) HV-based denoising; (c) NNV-based denoising}
 \label{heatmap}
\end{figure*}

This can be clearly observed from Fig. \ref{Fig.seg_dice_example}, where Fig. \ref{Fig.seg_dice_example} (b) is obtained by denoising dirty image (a) with RED-CNN guided by an HV rule. Visually, Fig. \ref{Fig.seg_dice_example} (b) has a much lower level of noise compared with Fig. \ref{Fig.seg_dice_example} (c), which is denoised by an NNV manner tailored for deep learning by deliberately keeping some noises. 
Yet surprisingly, processing both denoised images with the same segmentation network No-New-Net, suggests that the noisier one, i.e., Fig. \ref{Fig.seg_dice_example} (c) denoised by NNV, could achieve a much higher Dice score than the clean version, i.e., Fig. \ref{Fig.seg_dice_example} (b) denoised by HV on image segmentation. The detailed segmentation result comparison is illustrated in Fig. \ref{Fig.seg_dice_example} (e) and (f).

\section{The Difference between Human-Vision and Neural-Network-Vision}
\label{sec:difference}

\textbf{Frequency results}
In order to know the difference between HV and NNV, we first transfer an image into the frequency domain by $8\times8$ 2D Discrete Cosine Transform (DCT). In this way, the image is split into multiple small $8\times8$ frequency coefficients blocks. We then put the frequency coefficients belong to the same frequency components together to measure its distribution of this image (totally 64 frequency distributions). Since all the distributions obey normal distribution (i.e., mean is 0), thus the standard deviation (SD) indicates the energy in each frequency component of this image (i.e., large SD means more energy in this frequency component). 
Fig.~\ref{heatmap} shows the heat map of SD at each frequency component, where (a) is clean image, (b) indicates HV-based denoised image, and (c) represents the proposed NNV-based denoised image.

Obviously, the NNV-based denoised image (c) has more comprehensive information in high frequency domain compared with clean image (a). 



This indicates how the segmentation network wants to change the denoised image, i.e., the additional information added in NNV-based denoised image is guided by segmentation network.

\textbf{Frequency analysis}
Assume $x_k$ is a single pixel of a raw image \textbf{X}, and $x_k$ can be represented by $8\times8$ DCT: 
\begin{equation}
x_k=\sum_{i=0} ^{i=7}\sum_ {j=0} ^{n=7} c_{(k,i,j)}\cdot b_{(i,j)}
\end{equation}
where $c_{(k,i,j)}$ and $b_{(i,j)}$ are the DCT coefficient and corresponding basis function at 64 different frequencies, respectively.

Since the human visual system is less sensitive to high-frequency components, HV-based denoising is achieved by intentionally discarding the high-frequency parts $c_{(k,i,j)}$. On the contrary, Deep-Neural-Networks (DNN) examine the importance of the frequency information in a quite different way. The gradient of the DNN function $F$ with respect to a basis function $b_{(i,j)}$ can be calculated as:

\begin{equation}
{ \frac{\partial F}{\partial b_{(i,j)}}=\frac{\partial F}{\partial x_k}\times\frac{\partial x_k}{\partial b_{i,j}}=\frac{\partial F}{\partial x_k}\times c_{(k,i,j)} }
\label{basis}
\end{equation}

Eq.~\ref{basis} implies that the contribution of a frequency component ($b_{i,j}$) of a single pixel $x_k$ to the DNN learning will be mainly determined by its associated DCT coefficient ($c_{(k,i,j)}$) and the importance of the pixel ($\frac{\partial F}{\partial x_k}$). Here $\frac{\partial F}{\partial x_k}$ is obtained after the DNN training, while $c_{(k,i,j)}$ will be distorted by filtering before training. 
If $c_{(k,i,j)}=0$, the frequency feature ($b_{i,j}$), which may carry important details for DNN feature map extraction, cannot be learned by DNN for weights updating, causing a lower accuracy.

As shown in Fig~\ref{heatmap} (a), clean image has comprehensive information in all frequency domains, however (b) HV-based method discord the high frequency information which will make DNN hard to learn these features. The NNV-based method can add more features in all frequency components to make the DNN easier to learning or training.

\section{Neural-Network-Vision-based Denoising}
\label{sec:analysis}
In this section, we would like to discuss about the concept for the proposed framework. First, we would like to define the denoising network as function $f:R^{m\times n}\xrightarrow{}R^{m\times n}$ and the segmentation neural network as operation $g:R^{m\times n}\xrightarrow{}R^{k}$, where $m$ and $n$ are the width and the height of the input image, respectively and $k$ is the number of categories for each pixel.

For segmentation, the objective function of the training procedure which obtains the weights $\theta_{g}$ of model $g$ can be written as:
\begin{equation}
\begin{aligned}
    \min_{\theta_{g}} \mathcal{L}_{g}(g(x;\theta_{g}),y)
\end{aligned}
\end{equation}
where $x$ and $y$ are the input image and the ground-truth of the model, respectively. $\mathcal{L}_{g}$ denotes the loss function which is minimized by optimizer.

On the other hand, the HV-based denoising function $f$ is trained individually as:
\begin{equation}
\begin{aligned}
    \min_{\theta_{f}} \mathcal{L}_{f}(f(x;\theta_{f}),y)
\end{aligned}
\end{equation}
where the weights $\theta_{f}$ is optimized by minimizing loss function $\mathcal{L}_{f}$.

The proposed NNV denoising framework is proved to have at least same power with application model itself. The weights $\theta$ of the NNV-based denoising network inside proposed framework $f$ is minimized through loss function $\mathcal{L}_{g}$ as:

\begin{equation}
\begin{aligned}
    \min_{\theta_{f}} \mathcal{L}_{g}(g(f(x;\theta_{f});\theta_{g}),y)
\end{aligned}
\end{equation}
While $\theta_{f}$ is given in worst case, the denoising model $f$ could only learn to make $f(x;\theta_{f}) = x$, which means simply output the input. However, as substitute $f(x;\theta_{f}) = x$ into the framework $g(f(x;\theta_{f})$, it would just make the framework as $g(x;\theta_{g})$, which is equivalent with doing segmentation itself. Though this statement, we believe denoising network is required in our experiments.

\section{Experiments}
\label{sec:exp}
\subsection{Datasets}
We use two segmentation datasets and one classification dataset to evaluate the proposed framework. Start with the first segmentation dataset, Multi-Modality Whole Heart Segmentation (MM-WHS) \cite{zhuang2013challenges} dataset was acquired at Shanghai Shuguang Hospital, China, using routine cardiac CT angiography. All the image cover the whole heart from the upper abdominal to the aortic arch. The slices were acquired in the axial view. This dataset aims to accurately segment all the substructures of the whole heart into seven categories and background, as eight classes. In our experiment, we considered the dataset as clean images even though some of them still remain noises. We then synthesized the dirty dataset by superposing the noise to the clean image, which follows normal Gaussian distribution and Poisson distribution. Moreover, 2,557 images from 19 series are used as the training set, and the test set contains 363 images from 1 scanning series.

Second, we examined the experiments with Brain Tumor segmentation challenge (BraTS) \cite{menze2014multimodal} segmentation dataset.
This dataset includes 210 High Grade Glioma (HGG) cases, which consist of a T1 weighted, a post-contrast T1-weighted, a T2-weighted, and a Fluid-Attenuated Inversion Recovery (FLAIR) MRI for each patient. We chose post-contrast T1-weighted images as our input in the experiment. Each tumor is segmented into edema, necrosis, and non-enhancing tumor, and active/enhancing tumor, which is in 5 categories (background + 4 classes). We split the dataset into the training set and test set with 1,125 images and 129 images, respectively.

For classification dataset, we utilized a brain tumor public dataset \cite{Cheng2016brainTD}. The objective of this dataset is to correctly classify the input into one of the three grades (Grade II, Grade III, and Grade IV). This dataset contains 233 patients with a total of 3,064 brain images with meningioma, glioma, and pituitary tumors, which is in three grades. We split them into training set with 2,500 images and test set with 300 images, not including in the training set. The images are T1-weighted contrast enhanced MRI images of axial (transverse plane), coronal (frontal plane), or sagittal (lateral plane) planes.


\subsection{Experiment Schemes}
In this paper, we compared our Neural-Network-Vision (NNV) based denoising scheme with three other schemes, segmentation or classification network Trained with Clean images (TC) and Trained with Dirty images (TD) which has the same noise level as test images, and the Human-Vision (HV) based denoising, respectively.
To train all the four schemes, noises were added to datasets with different noise levels. For TC and TD schemes, they were trained using clean images and dirty images, respectively. For HV scheme, the denoising network was independently trained using paired clean images and dirty images and optimized using pixel-wise loss function, such as Mean Squared Error (MSE). Finally, the proposed framework, NNV scheme was trained with dirty images and optimized by the loss function which considered the difference between the output of the following neural networks and its ground-truth, such as cross-entropy loss.

\subsection{Experiment Setup}
In our paper, three application models were used, including a 3D segmentation model, a 2D segmentation model, and a 2D classification model. For 3D scenario, every experiment scheme involved required input volumes which were scaled into 64$\times$64$\times$64. As for the experiments based on the 2D model, the input images were scaled to the size of 256$\times$256. 
While training, each scheme required different training epoch. We followed the default settings mentioned in the referred paper. For HV and MV schemes, the number of epoch was set to 300. Xavier uniform initializer \cite{xavier2010understanding} was used for all kernel in every convolution and deconvolution layer. The batch size was set to 1. Adam optimization \cite{diederik2014adam} was used to minimize the loss function with learning rate at 0.00001. Moreover, all experiments done in our paper were implemented in Python3 with TensorFlow 1.14 over NVIDIA GeForce RTX 2080 Ti GPU.

\subsection{Referenced Application Neural Network Model}
In this section, we will briefly introduce the model we applied as the vehicle.

For denoising network, RED-CNN \cite{chen2017low} and MCDnCNN \cite{jiang2018denoising} were used in baseline framework and the proposed framework.
RED-CNN, known as Residual Encoder-Decoder Convolutional Neural Network, contains five convolution and deconvolution layers. With connection between encoder and decoder block, residual information could be carried to the latter layer. In Chen et al. \cite{chen2017low}, it is proposed to reconstruct a denoised image from low-dose CT image.
Multi-Channel Denoising Convolutional Neural Network (MCDnCNN) \cite{jiang2018denoising} is a modified version from DnCNN \cite{zhang2017beyond}, which considers the neighboring slices for a better result. It is basically formed by eight convolution layers with batch normalization and a final convolution layer. Proposed by Jiang et al., it aims to denoise noises inside MRI images.

\begin{figure}[htbp]
\begin{center}
\includegraphics[width=0.95\linewidth]{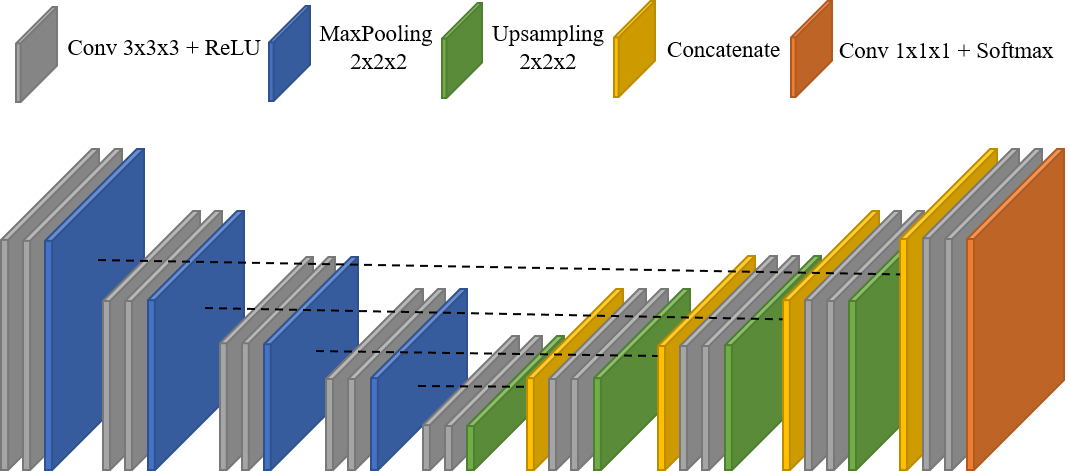}
\caption{The network structure of No-New-Net \cite{isensee2018no}, which is a famous U-Net based architecture.}
\label{Fig.U-Net}
\end{center}
\end{figure}

For segmentation, since U-Net based architecture is well-known for image segmentation, No-New-Net \cite{isensee2018no} 3D and 2D version were selected and implemented. Originally, No-New-Net was examined by using a brain tumor segmentation dataset, and results in rank two of brain tumor segmentation challenge (BraTS) 2018. Thus, we designed a similar model by replacing all 3D layers into 2D version. 

At last, we implemented the Classification Convolutional Neural Network (CCNN), which follows the model mentioned in Sultan et al. \cite{sultan2019multi}. The model is built based on basic convolutional neural network (CNN) with four max pooling layers. It is proposed to classify different grades of brain tumor through MRI image.

\begin{figure}[htbp]
\begin{center}
\includegraphics[width=0.95\linewidth]{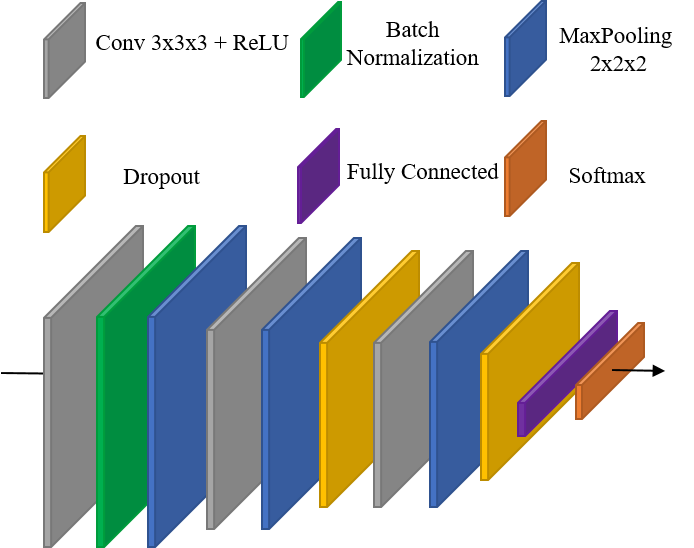}
\caption{The network structure of Classification Convolutional Neural Network (CCNN) \cite{sultan2019multi}.}
\label{Fig.Classification}
\end{center}
\end{figure}

\subsection{Evaluation Metrics}

For segmentation results, we follow existing works \cite{isensee2018no,mortazi2017multi,zhao2019data} applying Dice score and Hausdorff Distance $d_{H}$ for evaluation. Dice score is influenced more with the overlap percentage between prediction and ground-truth. Besides, Hausdorff Distance calculates the largest distance between prediction and ground-truth boundary, which is influenced by the boundary distance.

The two metrics could be formulated as:
\begin{equation}
\begin{aligned}
Dice = \frac{2 |A \cap B|}{|A|+|B|}
\end{aligned}
\end{equation}

\begin{equation}
\begin{aligned}
d_{H} = \max\left\{ \adjustlimits \sup_{a\in A} \inf_{b\in B} d(a,b), \adjustlimits \sup_{b\in B} \inf_{a\in A}d(a,b)\right\}
\label{eq:hausdorff}
\end{aligned}
\end{equation}

where A and B denote two sets as ground-truth and segmentation prediction, respectively. In Eq. \ref{eq:hausdorff}, $\sup$ and $\inf$ are supremum and infimum of sets, respectively.  $d(\cdot)$ could be defined as any distance calculation function. In our experiment, we applied Euclidean distance in our experiment. Moreover, most existing works \cite{karimi2019reducing} show that in medical imaging, a Dice improvement over 0.01 is already significant when comparing with the same neural networks with different settings.

For classification, top-k accuracy is the most common metric for evaluation. In our paper, since only a few categories were desired to be classified, top-1 accuracy was applied and reported in Section \ref{sec:results}.

\section{Results}
\label{sec:results}
In this section, we will discuss the experimental results which are completed using two datasets, two noise types, and two denoising networks on segmentation and classification for the four experiment schemes, as TC, TD, HV, and NNV schemes mentioned in Section \ref{sec:exp}. In order to show the flexibility of the proposed framework, we test all four trained schemes with different noise level included in the training set. 

\subsection{Results Analysis for Segmentation}

\begin{table*}[htbp]
\centering
\caption{Statistic result comparison (mean$\pm$SD) for No-New-Net 2D (256$\times$256) model using MM-WHS and BraTS datasets. RED-CNN and MCDnCNN denoising networks were used. Gaussian white noise $\sigma=70$ was added to simulate dirty images.}
\label{Table:seg}
\setlength{\tabcolsep}{4pt}
\begin{tabular}{cc|cc|cc|cc|cc}
\hline
 &  & \multicolumn{4}{c|}{Gaussian Noise} & \multicolumn{4}{c}{Poisson Noise} \\ \hline
 &  & \multicolumn{2}{c|}{MM-WHS dataset} & \multicolumn{2}{c|}{BraTS dataset} & \multicolumn{2}{c|}{MM-WHS dataset} & \multicolumn{2}{c}{BraTS dataset} \\ \hline
Schemes &  & Dice & Hausdorff & Dice & Hausdorff & Dice & Hausdorff & Dice & Hausdorff \\ \hline
\multirow{2}{*}{w/o Denoise} & TC & 0.542$\pm$0.251 & 2.723$\pm$0.976 & 0.481$\pm$0.184 & 3.047$\pm$0.837 & 0.641$\pm$0.217 & 2.655$\pm$1.725 & 0.525$\pm$0.186 & 2.502$\pm$1.055 \\
                             & TD & 0.676$\pm$0.257 & 2.815$\pm$2.116 & 0.497$\pm$0.203 & 2.570$\pm$0.760 & 0.703$\pm$0.160 & 2.685$\pm$1.538 & 0.588$\pm$0.168 & 2.334$\pm$0.732 \\ \hline
\multirow{2}{*}{RED-CNN}     & HV & 0.779$\pm$0.177 & 1.953$\pm$1.373 & 0.577$\pm$0.169 & 2.366$\pm$0.727 & 0.779$\pm$0.182 & 1.994$\pm$1.398 & 0.590$\pm$0.169 & 2.320$\pm$0.721 \\
                             & NNV & \textbf{0.798$\pm$0.167} & \textbf{1.899$\pm$1.333} & \textbf{0.585$\pm$0.164} & \textbf{2.340$\pm$0.781} & \textbf{0.829$\pm$0.150} & \textbf{1.790$\pm$1.227} & \textbf{0.602$\pm$0.163} & \textbf{2.253$\pm$0.740} \\ \hline
\multirow{2}{*}{MCDnCNN}     & HV & 0.783$\pm$0.176 & 1.965$\pm$1.392 & 0.575$\pm$0.172 & 2.416$\pm$0.745 & 0.757$\pm$0.190 & 2.067$\pm$1.419 & 0.596$\pm$0.161 & 2.326$\pm$0.699\\
                             & NNV & \textbf{0.802$\pm$0.171} & \textbf{1.896$\pm$1.340} & \textbf{0.585$\pm$0.163} & \textbf{2.341$\pm$0.774} & \textbf{0.820$\pm$0.155} & \textbf{1.850$\pm$1.293} & \textbf{0.600$\pm$0.169} & \textbf{2.281$\pm$0.770}\\ \hline
\end{tabular}
\end{table*}

We first evaluated how NNV can improve the segmentation accuracy over HV and segmentation network itself by using No-New-Net \cite{isensee2018no} 2D segmentation network. To show the feasibility, two datasets were applied to the experiment.

Table \ref{Table:seg} reports the mean and the standard deviation (SD) of the test result using MM-WHS and BrasTS datasets, with two different noise added, Gaussian noise and Poisson noise for four experiment schemes.
To show the flexibility of the proposed framework, two denoise methods, RED-CNN and MCDnCNN were implemented to both HV and NNV schemes. For a fair comparison, both schemes in each denoising method were trained with the same hyperparameters and settings.

\begin{figure}[]
\centering
\subfigure[Input]{
\includegraphics[width=0.25\linewidth]{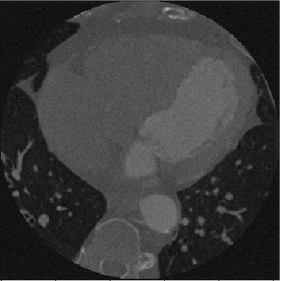}}
\subfigure[TC]{
\includegraphics[width=0.25\linewidth]{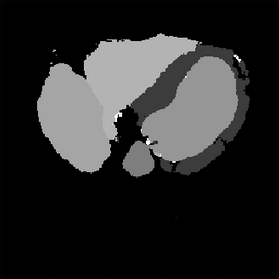}}
\subfigure[TD]{
\includegraphics[width=0.25\linewidth]{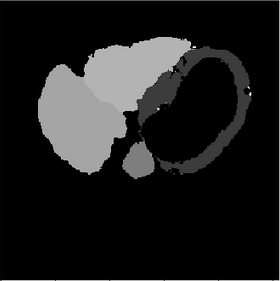}}
\subfigure[Ground-truth]{
\includegraphics[width=0.25\linewidth]{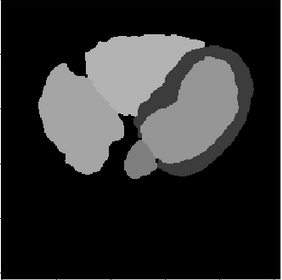}}
\subfigure[HV]{
\includegraphics[width=0.25\linewidth]{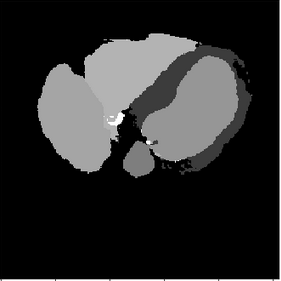}}
\subfigure[NNV]{
\includegraphics[width=0.25\linewidth]{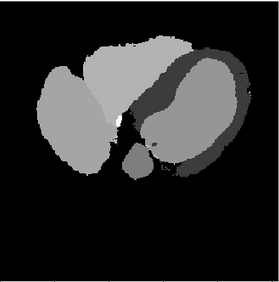}}
\caption{Segmentation comparison for (a) input image from MM-WHS dataset. (b-c) are the segmentation results of TC and TD schemes, which denoising network is not included. (d) is the ground-truth and (e-f) are the segmentation results of HV and NNV schemes using RED-CNN denoising network. Note that Gaussian white noise $\sigma=70$ is added to simulate dirty images in this case.}
\label{Fig.seg_mmwhs}
\end{figure}

We start our discussion on No-New-Net 2D segmentation. For MM-WHS dataset, first of all, TD scheme achieved 0.134 better Dice than TC scheme on average. This is as expected since the network could learn the feature extracted from noisier images. Secondary, since denoising method was included, we believe that HV denoise network is still somehow effective. Thus, for both denoising methods, RED-CNN and MCDnCNN, HV scheme improved the Dice and the Hausdorff over both schemes without denoising network by 0.103 and 0.107, 0.77 and 0.85, respectively. Finally, NNV scheme achieved another improvement over HV scheme by 0.019 higher Dice and 0.054 lower Hausdorff distance for RED-CNN and 0.018 and 0.069 for MCDnCNN. Fig. \ref{Fig.seg_mmwhs} shows the input, ground-truth, and the segmentation results of four schemes. 

As for the comparison with Poisson noise added in MM-WHS dataset, we can notice that TD scheme achieves higher Dice than TC scheme with 0.062 improvement. Compared with the TD scheme, the proposed NNV scheme achieves the optimal performance in both metrics which yields an improvement of 0.126 higher Dice and 0.895 lower Hausdorff distance for RED-CNN and 0.117 and 0.835 for MCDnCNN.

We also applied the same experiment to BraTS dataset. The statistic result of the experiment is also reported in Table \ref{Table:seg}. We can notice that the improvement trend of all the schemes on MM-WHS dataset is the same as that on BraTS dataset. As for Gaussian noise experiment, for those schemes without denoising, TD scheme outperforms TC scheme with Dice 0.016 on average. However, as the denoising network is included, HV scheme beats TD scheme with 0.08 and 0.204 on Dice score and Hausdorff distance, respectively. Finally, NNV scheme successfully results in the highest Dice 0.585 and the lowest Hausdorff distance 2.340 on average among all four schemes.
Poisson noise experiment also has a similar improvement trend compared with Gaussian noise experiment. TD scheme again improves the Dice score with 0.063 than TC scheme. HV scheme beat TD scheme with slightly improvement. And finally the proposed NNV scheme lead the optimal score in both Dice score and Hausdorff distance with 0.01 and 0.067, respectively for RED-CNN and 0.004 and 0.045, respectively for MCDnCNN.

We further applied the experiment to No-New-Net 3D version, the comparison are reported in Table \ref{Table:seg3d_mmwhs}. In this experiment, Gaussian white noise with $\sigma=90$ was superposed to the MM-WHS dataset. From the table, similar to 2D mmwhs segmentation experiment, TD scheme again achieves 0.01 higher Dice over the TC scheme. With denoising network included, HV and MV schemes outperform TD scheme up to 0.014 Dice and 0.019 sensitivity on average. Furthermore, compared with HV scheme, MV scheme achieves slightly higher Dice and sensitivity at 0.01 and 0.005, respectively for RED-CNN, and 0.004 and 0.001, respectively for MCDnCNN. However, since the size of input volume is only 64$\times$64$\times$64, all the four schemes had originally achieved over 0.8 Dice score. Thus, the improvement, though smaller than that in the No-New-Net 2D segmentation presented in the paper, is still significant. Moreover, it can be clearly seen that all the experiments result in high specificity, which means that most true negative cases can be correctly segmented.

\begin{table}[]
\centering
\caption{Statistic result comparison (mean$\pm$SD) for No-New-Net 3D (64$\times$64$\times$64) model using MM-WHS dataset. Both training and test set contains Gaussian white noise with $\sigma=90$.}
\label{Table:seg3d_mmwhs}
\begin{tabular}{cc|ccc}
\hline
Schemes                      &     & Dice                     & Sensitivity              & Specificity     \\ \hline
\multirow{2}{*}{w/o Denoise} & TC  & 0.817$\pm$0.089          & 0.817$\pm$0.093          & 0.997$\pm$0.001 \\
                             & TD  & 0.827$\pm$0.065          & 0.807$\pm$0.083          & 0.998$\pm$0.000 \\ \hline
\multirow{2}{*}{RED-CNN}     & HV  & 0.830$\pm$0.060          & 0.820$\pm$0.065          & 0.998$\pm$0.000 \\
                             & NNV & \textbf{0.840$\pm$0.053} & \textbf{0.825$\pm$0.061} & 0.998$\pm$0.000 \\ \hline
\multirow{2}{*}{MCDnCNN}     & HV  & 0.837$\pm$0.058          & 0.825$\pm$0.064          & 0.998$\pm$0.000 \\
                             & NNV & \textbf{0.841$\pm$0.054} & \textbf{0.826$\pm$0.062} & 0.998$\pm$0.000 \\ \hline
\end{tabular}
\end{table}

\subsection{Results Analysis for Classification}

\begin{table}[htbp]
\centering
\caption{Classification accuracy comparison for Classification Convolutional Neural Network (CCNN), using Brain Tumor dataset with three different noise levels $\sigma=50, 70, 90$ Gaussian white noise added. Note that train set and test set contain same noise level in this experiment.}
\label{Table:class_gaussian}
\begin{tabular}{c|c|c|c|c}
\hline
                                 & \multicolumn{2}{c|}{w/o Denoise} & \multicolumn{2}{c}{RED-CNN} \\ \hline
Cases                            & TC & TD & HV    & NNV \\ \hline
Gaussian white noise $\sigma=50$ & 0.370 & 0.936 & 0.946 & \textbf{0.960} \\ \hline
Gaussian white noise $\sigma=70$ & 0.350 & 0.923 & 0.933 & \textbf{0.940} \\ \hline
Gaussian white noise $\sigma=90$ & 0.343 & 0.890 & 0.926 & \textbf{0.936} \\ \hline
\end{tabular}
\end{table}

In this section, we also extend the evaluation to classification using CCNN proposed by Saltan et al. \cite{sultan2019multi}. Similar to previous section, we explored how the NNV-based denoising framework works among different noise levels. Thus, three datasets were synthesized with $\sigma = 50, 70, 90$ Gaussian white noise superposed to Brain Tumor dataset. Note that both training set and test set contain the same noise level.

From Table \ref{Table:class_gaussian}, we can again observe that in both TC and TD schemes, the higher level the noise is, the lower accuracy does the classification network results in. Obviously, TD scheme outperformed TC scheme in the experiment, which had up to 2.6$\times$ higher accuracy.
However, when it comes to comparison between HV and NNV schemes, all numbers outperformed schemes without denoising network. That is, denoising network is required for testing on dirty images. Based on the observation between HV and NNV scheme, NNV scheme successfully improved the accuracy up to 0.140 in all three cases.


\begin{figure}
    \centering
    \subfigure[Input]{
    \includegraphics[width=0.25\linewidth]{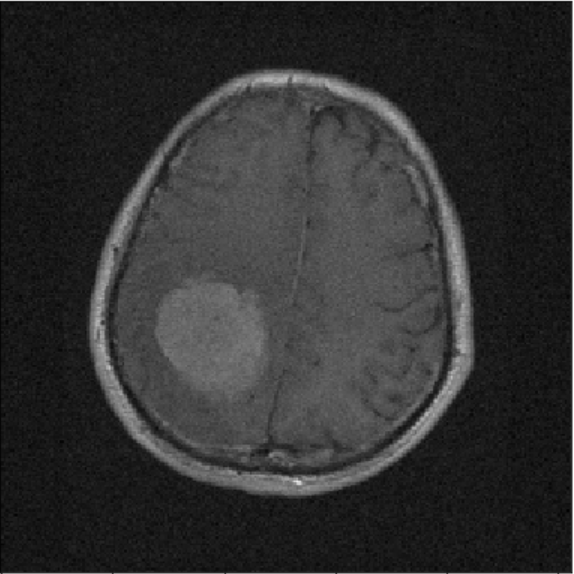}}
    \subfigure[HV scheme]{
    \includegraphics[width=0.25\linewidth]{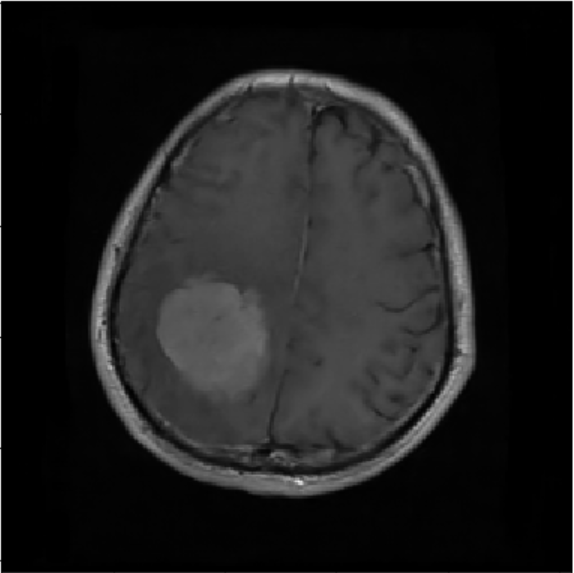}}
    \subfigure[NNV scheme]{
    \includegraphics[width=0.25\linewidth]{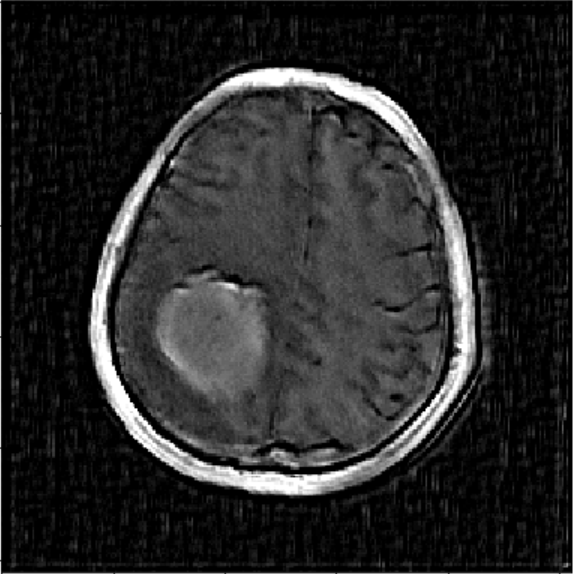}}
    \caption{Classification example for the (a) dirty input and the corresponded denoised image using (b) HV and (c) NNV experiment scheme from brain tumor dataset. The HV scheme misclassified the glioma into Grade III, which NNV scheme correctly classified into the Grade II.}
    \label{Fig.class_result}
\end{figure}

Furthermore, we also show a visualize example for classification in Fig. \ref{Fig.class_result}. In this case, glioma in Fig. \ref{Fig.class_result} (b) HV scheme was misclassified into Grade III. However, Fig. \ref{Fig.class_result} (c) NNV scheme successfully matched the correct class as Grade II. Since the confidence for Grade III and Grade II in this case is relatively close, as we sharpen the contour of the glioma become sharper in Fig \ref{Fig.class_result} (c), the classification model could classify the grade correctly.

\subsection{Results for Different Noise Levels in Training Set and Test Set}

\begin{table*}[]
\centering
\caption{Results (mean$\pm$SD) for different noise level occurred in training set and test set. Note that the experiment was trained using No-New-Net 2D segmentation on MM-WHS and BraTS dataset, respectively with $\sigma=70$ Gaussian white noise added.}
\label{Table:noise_level_seg_mmwhs}
\begin{tabular}{c|c|c|c|c|c|c}
\hline
\multirow{2}{*}{Dataset}                       & \multirow{2}{*}{Noise Level}&           & \multicolumn{2}{c|}{w/o Denoise}           & \multicolumn{2}{c}{RED-CNN}                \\ \cline{3-7}
                                               &                             &           & TC              & TD                       & HV              & NNV                      \\ \hline
\multirow{4}{*}{\specialcell{MM-WHS\\dataset}} &\multirow{2}{*}{$\sigma=50$} & Dice      & 0.574$\pm$0.247 & 0.684$\pm$0.259          & 0.717$\pm$0.191 & \textbf{0.721$\pm$0.192} \\ 
                                               &                             & Hausdorff & 2.583$\pm$1.005 & 2.789$\pm$2.127          & 2.168$\pm$1.456 & \textbf{2.132$\pm$1.398} \\ \cline{2-7}
                                               &\multirow{2}{*}{$\sigma=150$}& Dice      & 0.396$\pm$0.229 & \textbf{0.605$\pm$0.209} & 0.455$\pm$0.243 & 0.488$\pm$0.231          \\ 
                                               &                             & Hausdorff & 3.411$\pm$0.966 & 3.150$\pm$1.950          & 3.076$\pm$0.998 & \textbf{3.011$\pm$0.989} \\ \hline
\multirow{4}{*}{\specialcell{BraTS\\dataset}}  &\multirow{2}{*}{$\sigma=50$} & Dice      & 0.529$\pm$0.192 & 0.521$\pm$0.205          & 0.565$\pm$0.174 & \textbf{0.573$\pm$0.177} \\ 
                                               &                             & Hausdorff & 2.605$\pm$0.653 & 2.461$\pm$0.804          & 2.420$\pm$0.683 & \textbf{2.352$\pm$0.808} \\ \cline{2-7}
                                               &\multirow{2}{*}{$\sigma=150$}& Dice      & 0.437$\pm$0.178 & 0.504$\pm$0.180          & 0.580$\pm$0.160 & \textbf{0.584$\pm$0.170} \\ 
                                               &                             & Hausdorff & 3.406$\pm$0.920 & 2.492$\pm$0.749          & 2.405$\pm$0.707 & \textbf{2.333$\pm$0.778} \\ \hline
\end{tabular}
\end{table*}

\begin{table}[]
\centering
\caption{Results for different noise level occurred in training set and test set. Note that all experiment is trained for classification using brain tumor dataset with $\mu=0$, $\sigma=70$ Gaussian noise added.}
\begin{tabular}{c|c|c|c|c}
\hline
                                 & \multicolumn{2}{c|}{w/o Denoise} & \multicolumn{2}{c}{RED-CNN} \\ \hline
Cases                            & TC     & TD                      & HV    & NNV                 \\ \hline
Gaussian white noise $\sigma=50$ & 0.370  & 0.860                   & 0.953 & \textbf{0.973}      \\ \hline
Gaussian white noise $\sigma=90$ & 0.343  & 0.906                   & 0.920 & \textbf{0.923}      \\ \hline
\end{tabular}
\label{Table:noise_level_class}
\end{table}

In this section, we test all the four trained schemes with different noise levels from the training set to show the effectiveness of our framework. 

Firstly, we trained the four schemes for No-New-Net 2D segmentation network using MM-WHS and BraTS datasets with $\sigma = 70$ Gaussian white noise added. Table \ref{Table:noise_level_seg_mmwhs} shows the test results for test set with $\sigma = 50$ and $\sigma = 150$ Gaussian white noise superposed for MM-WHS and BraTS datasets, respectively. For both datasets, our expectation still holds. The two schemes with denoising network included outperformed the other two without denoising up to 0.037 higher Dice and 0.657 Hausdorff distance for MM-WHS dataset. For BraTS dataset, similar result obtained. NNV scheme leaded the best Dice 0.584 and Hausdorff distance 2.333 among all four experiment schemes.

Second, we brought our experiment to classification. In Table \ref{Table:noise_level_class}, we applied the four schemes, which were trained with images containing Gaussian white noise with $\sigma = 70$ to test set with Gaussian white noise $\sigma = 50$ and $\sigma = 90$, respectively. Obviously, all the four schemes performed as our expectation. NNV scheme outperformed the other three schemes in all the three cases, where HV, TD, and TC schemes had 0.02, 0.11, and 0.6 higher accuracy, respectively for Gaussian white noise with $\sigma = 50$ and 0.003, 0.017, and 0.58, respectively for Gaussian white noise with $\sigma = 90$.

\section{Conclusion}
\label{sec:conclusion}

In this paper, due to the observation that neural network applications focus on different sight from human eyes, we introduced a neural-network-vision-based denoising framework. Unlike previous human-vision-based denoising methods, our framework could perform a better result for neural network application.
By evaluating the experiment through different networks, noise types, and datasets on segmentation and classification, experimental results have shown the effectiveness and the feasibility of the proposed framework.



\ifCLASSOPTIONcaptionsoff
  \newpage
\fi


\end{document}